\documentclass[%
 reprint,
superscriptaddress,
 amsmath,amssymb,
 aps,
pra,
floatfix,
]{revtex4-2}

\usepackage{graphicx}
\usepackage{xcolor}
\usepackage{dcolumn}
\usepackage{bm}
\usepackage{tensor}
\usepackage{gensymb}
\usepackage{xr-hyper} 
\usepackage{hyperref}

\makeatletter

\newcommand*{\addFileDependency}[1]{
\typeout{(#1)}
\@addtofilelist{#1}
\IfFileExists{#1}{}{\typeout{No file #1.}}
}\makeatother

\newcommand*{\myexternaldocument}[1]{%
\externaldocument{#1}%
\addFileDependency{#1.tex}%
\addFileDependency{#1.aux}%
}
\myexternaldocument{supp}

\begin{document}

\preprint{APS/123-QED}

\title{Optical Telecommunications-Band Clock based on Neutral Titanium Atoms}

\author{Scott Eustice}%
    \affiliation{Department of Physics, University of California, Berkeley, CA 94720}
    \affiliation{Challenge Institute for Quantum Computation, University of California, Berkeley, CA 94720}
\author{Dmytro Filin}
    \affiliation{Department of Physics and Astronomy, University of Delaware, Newark, DE 19716}
\author{Jackson Schrott}
    \affiliation{Department of Physics, University of California, Berkeley, CA 94720}
    \affiliation{Challenge Institute for Quantum Computation, University of California, Berkeley, CA 94720}
\author{Sergey Porsev}
    \affiliation{Department of Physics and Astronomy, University of Delaware, Newark, DE 19716}
\author{Charles Cheung}
    \affiliation{Department of Physics and Astronomy, University of Delaware, Newark, DE 19716}
\author{Diego Novoa}
    \affiliation{Department of Physics, University of California, Berkeley, CA 94720}
    \affiliation{Challenge Institute for Quantum Computation, University of California, Berkeley, CA 94720}
\author{Dan M. Stamper-Kurn}
    \affiliation{Department of Physics, University of California, Berkeley, CA 94720}
    \affiliation{Challenge Institute for Quantum Computation, University of California, Berkeley, CA  94720}
    \affiliation{Materials Science Division, Lawrence Berkeley National Laboratory, Berkeley, CA 94720}
\author{Marianna S. Safronova}
    \affiliation{Department of Physics and Astronomy, University of Delaware, Newark, DE 19716}
    \affiliation{Joint Quantum Institute, National Institute of Standards and Technology and the University of Maryland, College Park, Maryland 20742}

\date{\today}

\begin{abstract}
We propose an optical clock based on narrow, spin-forbidden M1 and E2 transitions in laser-cooled neutral titanium.  These transitions exhibit much smaller black body radiation shifts than those in alkaline earth atoms, small quadratic Zeeman shifts, and have wavelengths in the S, C, and L-bands of fiber-optic telecommunication standards, allowing for integration with robust laser technology.  We calculate lifetimes; transition matrix elements; dynamic scalar, vector, and tensor polarizabilities; and black body radiation shifts of the clock transitions using a high-precision relativistic hybrid method that combines a configuration interaction and coupled cluster approaches.  We also calculate the line strengths and branching ratios of the transitions used for laser cooling.  To identify magic trapping wavelengths, we have completed the largest-to-date direct dynamical polarizability calculations.  Finally, we identify new challenges that arise in precision measurements due to magnetic dipole-dipole interactions and describe an approach to overcome them.  Direct access to a telecommunications-band atomic frequency standard will aid the deployment of optical clock networks and clock comparisons over long distances.
\end{abstract}

\maketitle


Optical atomic clocks have taken a giant leap in recent years, with several experiments reaching uncertainties at the 10$^{-18}$ level~\cite{brewer__2019,sanner_optical_2019,bothwell_jila_2019}. The comparison of clocks based on different atomic standards~\cite{boulder_atomic_clock_optical_network_bacon_collaboration_frequency_2021} or placed in separate locations~\cite{barontini_measuring_2022} enables important applications such as relativistic geodesy~\cite{mcgrew_atomic_2018}, tests of fundamental physics~\cite{safronova_search_2018}, and dark matter searches~\cite{antypas_new_2022}.  These applications motivate the development of synchronized clock networks and transportable clocks that operate in extreme and distant environments~\cite{buchmueller_snowmass_2022}.

The leading neutral-atom optical clocks operate on wavelengths of 698 nm (Sr)~\cite{takamoto_optical_2005} and 578 nm (Yb)~\cite{lemke_spin-_2009}.  Light at these wavelengths is strongly attenuated in optical fibers, posing a challenge to long-distance time transfer.  These wavelengths are also inconvenient for constructing the ultrastable lasers that are an essential component of optical clocks.

By comparison, an optical atomic clock operating in the telecommunication wavelength band would have clear advantages.  The S-, C- and L-bands, ranging altogether between about 1460 and 1625 nm, feature low losses in standard optical fibers.  Stable light sources and robust optical amplifiers are also available across these ranges~\cite{winzer_fiber-optic_2018}.  These features would support the development of fiber-linked terrestrial clock networks over continental distances.

\begin{figure}[ht]
    \centering
    \includegraphics[width=.95\linewidth]{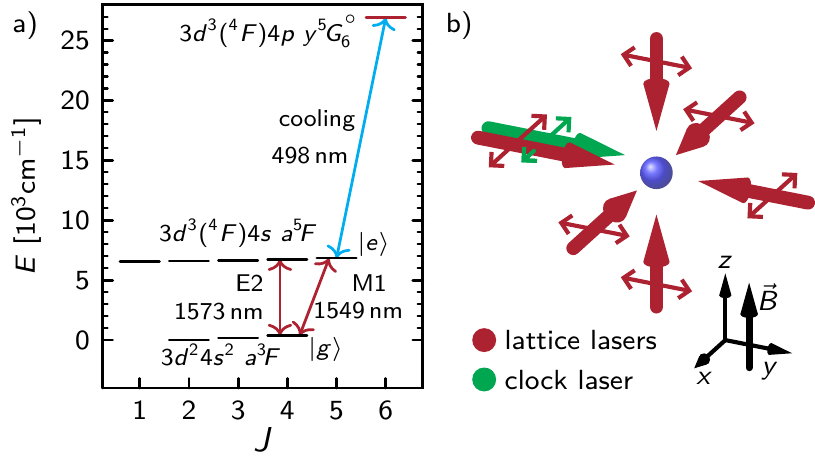}
    \caption{(a) Relevant atomic structure in Ti for an optical clock.  The $a\tensor*[^3]{F}{}$ and $a\tensor*[^5]{F}{}$ terms serve as the basis for the optical clock, while the excited $y\tensor*[^5]{G}{^\circ_6}$ level serves as the excited state for laser cooling of Ti.  The two optical clock transitions highlighted in the text are shown as maroon arrows; the laser cooling transition is shown in cyan.  (b) A diagram of the proposed experimental system.  Polarizations are indicated on a given beam by a small arrow of the same color as the beam itself.}
    \label{fig:ti_struc}
\end{figure}

We propose the use of ultra-narrow optical transitions in atomic titanium (Ti) as the basis of a telecommunications-band atomic clock.  It has recently been pointed out that numerous transition-metal elements, including Ti, can be laser-cooled on near-cycling optical transitions~\cite{eustice_laser_2020}, allowing for the adoption of optical lattice or tweezer trapping techniques~\cite{young_half-minute-scale_2020} used in today's leading neutral-atom clocks.  We identify several transitions between the $3d^24s^2$ $a\tensor*[^3]{F}{}$ and $3d^3(\tensor*[^4]{F}{})4s$ $a\tensor*[^5]{F}{}$ fine structure manifolds in Ti with transition wavelengths between 1483 and 1610 nm (see Fig.~\ref{fig:ti_struc} and Table~\ref{tab:clock_list}) that can serve as optical clock references for ultrastable telecommunication-band light sources.

From a numerical calculation of the Ti level structure, we identify several key features that make Ti an attractive atom for clock applications: the extreme narrowness of the candidate clock transitions, a weak clock sensitivity to blackbody radiation shifts, and the existence of several magic wavelengths for optical trapping.  While we identify challenges posed by the non-zero angular momentum of the clock states in Ti, we show that a proper magic-wavelength condition for optical trapping, which imposes a significant differential tensor ac Stark shift, mitigates their effects.

Our analysis relies on high precision atomic structure calculations, by which we characterize 85 levels of neutral Ti.  For this, we employ a hybrid method that combines the configuration interaction (CI) and linearized coupled-cluster (CC) approaches (referred to as CI + all order method~\cite{safronova_development_2009,norrgard_laser_2022}).  In this method, the correlations between four valence electrons are included via a large-scale CI computation using a highly parallel message passing interface (MPI) CI code~\cite{cheung_scalable_2021,norrgard_laser_2022}.  Several computations with increased number of configurations were carried out to ensure convergence.  The core-core and core-valence correlations are included using an effective Hamiltonian formalism~\cite{safronova_development_2009}.  We construct the effective Hamiltonian using second-order many-body perturbation theory (MBPT) and more accurate CC methods.  The difference between these results gives the size of the higher-order corrections, which we use to estimate uncertainties on all theory values~\cite{norrgard_laser_2022}.  The results are used to calculate transition rates, dynamical polarizabilities, and systematic shifts in the clock transitions.  Further details of the computational methods are given in the Supplemental Material~\cite{suppref}.

Several clock transitions are identified in Table~\ref{tab:clock_list}.  Transitions between the $a\tensor*[^3]{F}{}$ and $a\tensor*[^5]{F}{}$ manifolds occur via spin forbidden electric quadrupole (E2) and magnetic dipole (M1) transitions.  Calculated reduced matrix elements for these transitions are tabulated.  The calculated natural linewidths account for both the decay of the upper state to the lower manifold on the listed E2 and M1 transitions and the M1 decays within each fine-structure manifold.  The transitions are all exceptionally narrow, allowing for optical atomic clocks with long coherence times.

\begin{table}[ht]
    \centering
    \begin{ruledtabular}
    \begin{tabular}{ccccccc}
        $J$ & $J'$ & $\lambda$ (nm) & Tele. & $D_{\text{M1}}$ & $D_{\text{E2}}$ & $\Gamma$ \\
        &&& Band & ($10^{-3}\mu_B$) & (a.u.) & ($10^{-6}\text{s}^{-1}$)\\ \hline
        \bfseries 4 & \bfseries 5 & \bfseries 1548.926 & \bfseries C & \bfseries 1.0(5) & \bfseries 0.140(4) & \bfseries 242(5) \\
        \bfseries 4 & \bfseries 4 & \bfseries 1573.346 & \bfseries L & \bfseries 0.36(18) & \bfseries 0.134(8) & \bfseries 239(5) \\
        4 & 3 & 1593.846 & L & 1.02(12) & 0.0015(3) & 227(5) \\
        4 & 2 & 1609.816 & L & N/A & 0.0314(27) & 214(5) \\
        3 & 5 & 1498.615 & S & N/A & 0.0472(7) & 162.2(2.6) \\
        3 & 4 & 1521.463 & S & 0.4(4) & 0.027(10) & 159.1(2.5) \\
        3 & 3 & 1540.625 & C & 0.2(2) & 0.124(4) & 147.2(2.6) \\
        3 & 2 & 1555.541 & C & 0.3(4) & 0.0204(22) & 134.3(2.6) \\
        3 & 1 & 1565.754 & L & N/A & 0.0463(23) & 129.2(2.5) \\
        2 & 4 & 1483.073 & S & N/A & 0.0196(26) & 32.75(29) \\
        2 & 3 & 1501.275 & S & 0.40(16) & 0.024(7) & 20.83(36) \\
        2 & 2 & 1515.435 & S & 0.1(1) & 0.1006(24) & 7.93(38) \\
        2 & 1 & 1525.127 & S & 0.23(2) & 0.0643(11) & 2.85(11)
    \end{tabular}
    \end{ruledtabular}
    \caption{List of proposed optical clock transitions in Ti.  All transitions are between the lower $a\tensor*[^3]{F}{}$ and upper $a\tensor*[^5]{F}{}$ terms.  The lower (upper) states are indexed by $J$ ($J'$).  Transition wavelengths $\lambda$ are taken from Ref.~\cite{kramida_nist_1999}.  The telecomm band is indicated, with S (short), C (conventional) and L (long) bands noted.  M1, E2 reduced matrix elements $D_{\text{M1}}$, $D_{\text{E2}}$ and transition linewidths $\Gamma$ are calculated.  The two clock transitions highlighted in the text are in bold.}
    \label{tab:clock_list}
\end{table}

In this letter, we focus on the $a\tensor*[^3]{F}{_4} \rightarrow a\tensor*[^5]{F}{_5}$ transition at $1549$ nm unless otherwise noted.  An advantage of this transition is that the $a\tensor*[^5]{F}{_5}$ state is the lower level of the near-cycling 498 nm transition, which is suited for laser cooling.  Our calculations predict that the cooling transition has low branching ratios to other even parity states ($\sim10^{-6}$), enabling single-laser state preparation and readout for atoms in the upper clock state.  For details on calculations relevant to the laser cooling transition, see the Supplemental Material~\cite{suppref}.  An additional benefit is that light at the 1549 nm clock wavelength can be generated by narrow-linewidth, high-power Er-doped fiber lasers, simplifying the required optical setup.

We consider the three titanium isotopes which have zero nuclear spin, and therefore no hyperfine structure ($^{46, 48, 50}$Ti).  To make the clock insensitive to first-order differential Zeeman shifts from stray magnetic fields, we drive the $|m_J = 0\rangle \to |m'_J = 0 \rangle$ transition, with $m_J$ being the magnetic quantum number and the primed symbols and numbers referring to the upper $a\tensor*[^5]{F}{}$ state.  Because the E2 matrix element for this transition is zero, only the M1 matrix element contributes to a direct one-photon drive of the clock transition.  Choosing quantization, clock-laser polarization, and clock-laser propagation axes as shown in Fig.~\ref{fig:ti_struc}, we calculate that for a driving intensity of $0.1$ $\text{W}/\text{mm}^2$, we achieve a clock Rabi frequency of 91(46) Hz.

To compare the strength of this M1 transition to that of an E2 transition in the same set of transitions, we also consider driving the $|a\tensor*[^3]{F}{_4},m_J=0\rangle\to|a\tensor*[^5]{F}{_4},m'_J=0\rangle$ transition at a wavelength of 1573 nm.  For this transition, the M1 matrix element vanishes while the E2 matrix element does not.  With the same intensity and polarization as in Fig.~\ref{fig:ti_struc}, but propagating along the $z$ axis, the Rabi frequency for such an E2 transition is 214(13) Hz.  For a detailed derivation of these Rabi frequencies, see the Supplemental Material~\cite{suppref}.

Neutral-atom optical clocks often use optical lattice potentials to confine atoms, allowing for a long interrogation time.  In order to avoid imposing large differential ac Stark shifts between the upper and lower states of the clock transition, it is necessary to use lattice light which is at a ``magic wavelength'', at which the dynamic polarizabilities of the lower and upper clock states are identical~\cite{takamoto_spectroscopy_2003}.

\begin{figure}[ht]
    \centering
    \includegraphics[width=\linewidth]{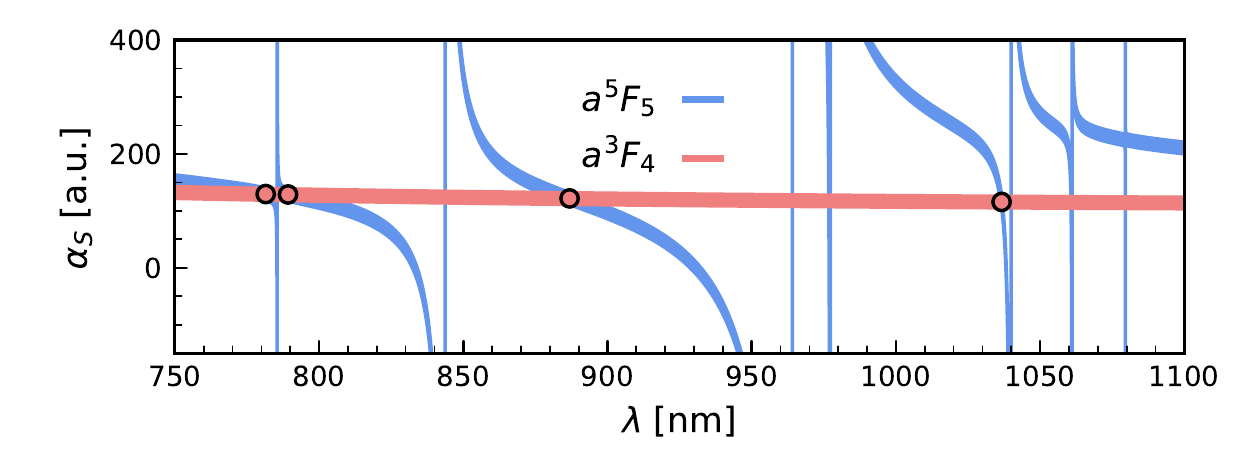}
    \caption{The scalar dynamic polarizability of the $m_J=0$ sublevels of the $a\tensor*[^3]{F}{_4}$ (red) and $a\tensor*[^5]{F}{_5}$ (blue) states in Ti from 1100 nm to 750 nm as calculated by the sum-over-states method.  The angle between the polarization and the $B$ field direction is set to 90\degree.  The locations of magic wavelengths considered in the rest of this work are circled.}
    \label{fig:magic_wavelengths}
\end{figure}

\begin{table*}[ht]
    \centering
    \begin{ruledtabular}
    \begin{tabular}{cccccccc}
        $\lambda_{\text{magic}}$ & $\alpha$ & $\alpha^S_{a\tensor*[^3]{F}{_4}}$ & $\alpha^S_{a\tensor*[^5]{F}{_5}}$ & $\alpha^V_{a\tensor*[^3]{F}{_4}}$ & $\alpha^V_{a\tensor*[^5]{F}{_5}}$ & $\alpha^T_{a\tensor*[^3]{F}{_4}}$ & $\alpha^T_{a\tensor*[^5]{F}{_5}}$ \\ \vspace{-3mm}\\\hline
        1036.6$^{+0.4}_{-0.4}$ & 116(10) &115(3) &66(12) & -2(2) & -470(30) & 4(4) & 154(8) \\
        887$^{+4}_{-4}$ & 122(5) &121(4) &159.3(2.6) & -3(2) & -104(5) & 5(4) & -111(3) \\
        789$^{+5}_{-2.2}$ & 129(5) &127(4) &127.1(1.4) & -4(3) & 127(4) & 5(4) & 6.5(1.6) \\
        781$^{+3}_{-7}$ & 130(5) &128(4) &126.3(1.3) & -4(3) & 138(3) & 6(4) & 11.5(1.5)
    \end{tabular}
    \caption{Data for the magic wavelengths for the $a\tensor*[^3]{F}{_4}$ to $a\tensor*[^5]{F}{_5}$ clock transition.  Wavelengths are given in units of nm, polarizabilities are given in atomic units.}
    \label{tab:magic_wavelengths_short}
    \end{ruledtabular}
\end{table*}

In Fig.~\ref{fig:magic_wavelengths} and Table~\ref{tab:magic_wavelengths_short}, we report several magic wavelengths for the $|a\tensor*[^3]{F}{_4}, m_J=0\rangle\to|a\tensor*[^5]{F}{_5}, m'_J = 0\rangle$ clock transition.  As with most states in Ti, the clock states experience significant vector and tensor ac Stark shifts~\cite{eustice_laser_2020}, owing to their non-zero angular momentum and Ti's complex spectrum.  To account for these shifts, we consider the specific lattice configuration shown in Fig.~\ref{fig:ti_struc}.  Here, a magnetic field applied in the $z$ direction imposes a linear Zeeman shift and defines the quantization direction.  All lattice light is linearly polarized in the transverse $x-y$ plane.  In this configuration, the clock transition is shifted only by the differential scalar and tensor ac Stark effects (the vector shift is zero on the $m_J=m'_J=0$ sublevels).  The sum of the scalar and tensor dynamic polarizabilities ($\alpha^S_i$ and $\alpha^T_i$ respectively) on the transition is then given by
\begin{equation}
    \Delta\alpha = \alpha^S_{a\tensor*[^5]{F}{_5}}-\alpha^S_{a\tensor*[^3]{F}{_4}}+\frac{1}{2}\left(\frac{2}{3}\alpha^T_{a\tensor*[^5]{F}{_5}}-\frac{5}{7}\alpha^T_{a\tensor*[^3]{F}{_4}}\right)
\end{equation}
At the identified magic wavelengths, the net transition ac Stark shift is zero.  For a more detailed description of the ac Stark shifts, see the Supplemental Material~\cite{suppref}.

Calculations of the polarizabilities were performed by two methods.  First, the sum-over-states method was used to roughly calculate polarizabilities over a wide range of frequencies.  The 76 transitions with the largest contributions to dc polarizability were used in the case of the $a\tensor*[^3]{F}{_4}$ states, while 51 transitions were used in the case of the $a\tensor*[^5]{F}{_5}$ states.  Once promising candidates for magic wavelengths were found, we performed direct dynamical polarizability calculations to identify the location of the magic wavelengths more precisely.  Direct computations for two of the magic wavelengths allow us to predict the remaining values accurately.  Previously, the direct computation method was only used for divalent systems such as Sr~\cite{safronova_blackbody-radiation_2013, kestler_magic_2022}, Mg~\cite{kulosa_towards_2015}, Yb~\cite{safronova_ytterbium_2012, tang_magic_2018}, Cd~\cite{yamaguchi_narrow-line_2019}, or Tm~\cite{golovizin_inner-shell_2019}.

For more complex atoms, the rapidly increasing number of relevant configurations makes such a direct computation intractable.  Here, we apply instead a truncation approximation: we order the configurations by weight to select the most important ones and then start removing configurations while checking the accuracy of the energies and relevant matrix elements.  This procedure drastically reduces the number of Slater determinants required to maintain numerical accuracy.  Further details on our method are found in the Supplemental Material~\cite{suppref}.  We emphasize that our approach is not specific to Ti; it should allow for the computation of polarizabilities, magic wavelengths, and other atomic properties for other atoms with a complex electronic structure.

Using the lattice configuration and magic wavelength described above not only eliminates the differential light shift, but also protects against the effects of dipole-dipole interactions between Ti atoms.  These effects are not present in lattice clocks of Sr, Yb, or Hg as those clocks operate on transitions between non-magnetic $J=0$ states.  In contrast, the magnetic moments of the proposed Ti clock states are both large, with $\mu_{a\tensor*[^3]{F}{_4}}=5.00\, \mu_B$ and $\mu_{a\tensor*[^5]{F}{_5}}=7.05\, \mu_B$.

There are three processes associated with the dipole-dipole interaction that we consider: dipolar relaxation, elastic spin-spin energy shifts, and inelastic spin-spin mixing~\cite{chomaz_dipolar_2022}.  Dipolar relaxation is the process by which Zeeman energy is converted to kinetic energy, depleting atoms from the clock states.  Such relaxation can be suppressed for atoms trapped in a deep 3D optical lattice by ensuring the bandgap far exceeds the Zeeman energy~\cite{de_paz_resonant_2013}.  The band energy scale in a lattice is set by the lattice recoil energy $E_r = h^2 / (8 m a^2)$, where $a$ is the lattice spacing.  For a $^{48}$Ti atom in the magic-wavelength lattice described above, the recoil energy is $E_r = h \times 6.8$ kHz.  3D optical lattice clocks typically use deep lattices to suppress tunneling and atom-atom contact interactions. As of 2019, the fermionic Sr 3D lattice clock at JILA operated at a lattice depth of $V_0=80E_r$~\cite{campbell_fermi-degenerate_2017}.  In deep lattices, the gap above the ground band is $E_g\approx2\sqrt{V_0E_r}$.  In the case of Ti, a comparable lattice operating at the magic wavelength near 781 nm could be achieved by intersecting six 3.5 W beams with waists of 0.1 mm.  This would give a lattice depth of $V_0\approx79E_r=h \times 540$ kHz and band gap of $E_g\approx18E_r=h \times 120$ kHz.  Setting the Zeeman energy below this band gap requires the ambient magnetic field be well below $B \sim E_g/\mu_B = 60$ mG.

The second two processes associated with the dipole-dipole interaction are captured in the so-called secular Hamiltonian, which is obtained by time-averaging the dipole-dipole interaction over Larmor precession:
\begin{multline}
    H_{dd}=\frac{\mu_0\mu_B^2}{8\pi}\sum_{\langle i,j\rangle}\frac{g_{J_i}g_{J_j}}{r_{ij}^3}\left(1-3\cos^2\theta_{ij}\right)\\
    \times\left(J^z_iJ^z_j-\frac{1}{4}\left(J^+_iJ^-_j + J^-_iJ^+_j\right)\right),
\end{multline}
Here, $i$ and $j$ label two atoms held at different sites of a lattice, separated by a distance vector of length $r_{ij}$ and polar angle $\theta_{ij}$ with respect to the quantization axis. $g_{J_i}$ is the Land\'{e} $g$-factor of the atom at lattice site $i$.

The elastic spin-spin energy shift corresponds to the $J_i^z J_j^z$ term in the secular Hamiltonian.  In theory, this term generates shifts to the transition frequency between atomic states with non-zero angular momenta.  However, for a clock transition between $m_J = m'_{J} = 0$ magnetic sublevels, the shift is zero and can be ignored.

The final process is the spin-mixing interaction, which corresponds to the $J^+_iJ^-_j + J^-_iJ^+_j$ term in the Hamiltonian.  This term couples atoms in an initial two-body state $|m^{(1)}_J = 0, m^{(2)}_J=0\rangle$ to final states $|m^{(1)}_J = \pm n, m^{(2)}_J = \mp n\rangle$, $n\in\{1,\dots,J\}$.  If not controlled, this would lead to rapid loss of population from the $m_J=0$ clock states.  The maximal strength of the coupling is $\hbar\Omega_{SM}=\mu_0 \mu_B^2 g_J^2 J(J+1) \sqrt{2} / 16 \pi (\lambda /2)^3$.  In a $\lambda=781$ nm optical lattice, this gives spin mixing strengths of $ h \times 2.4$ Hz (4.6 Hz) within the lower (upper) clock state manifold. Spin mixing between atoms in the upper and lower clock states is energetically suppressed because of the significant differential Zeeman splitting. For a 30 mG magnetic field, the splitting between the $|m^{(1)}_J = 0, m'^{(2)}_J=0\rangle$ and $|m^{(1)}_J = \pm1, m'^{(2)}_J = \mp1\rangle$ states is $h\times6.7$ kHz.

In the case where both atoms occupy either the upper or lower clock state, spin mixing is suppressed by the tensor ac Stark shift imparted by the optical lattice light. The tensor light shift creates an energy splitting between the $|m^{(1)}_J = 0, m^{(2)}_J=0\rangle$ and $|m^{(1)}_J = \pm n, m^{(2)}_J = \mp n\rangle$ two-atom states. Using the same optical lattice configuration described above, the splitting between the $|m^{(1)}_J = 0, m^{(2)}_J=0\rangle$ and $|m^{(1)}_J = \pm1, m^{(2)}_J=\mp1\rangle$ states is $\Delta E_{tens}=h \times 4(2)$ kHz ($h \times 4.8(6)$ kHz) within the lower (upper) clock state manifold.  Since the differential Zeeman splitting and $\Delta E_{tens}$ are much larger than $\hbar\Omega_{SM}$, spin mixing is highly suppressed.  

In this regime, spin mixing enters as a second-order perturbative effect.  The $|m^{(1)}_J = 0, m^{(2)}_J=0\rangle$ two-atom states in both the lower and upper clock manifolds are weakly coupled to the corresponding $|m^{(1)}_J = \pm1, m^{(2)}_J=\mp1\rangle$ states by $\Omega_{SM}$.  Both clock states experience an energy shift on the order of $\sim\Omega^2_{SM}/\Delta E_{tens}$.  The difference between the shifts leads to a shift of the clock frequency, while the sum of the shifts leads to decoherence between the clock states.  For two atoms, the shift is $\sim3$ mHz and the rate of decoherence is $\sim6$ mHz.  For more discussion of the dipole-dipole interaction, see the Supplemental Materials~\cite{suppref}.

One complication in our scheme of using tensor light shifts to combat magnetic dipole-dipole interactions is that deviations from the lattice-light polarization shown in Fig.~\ref{fig:ti_struc} will introduce clock frequency shifts.  Considering the example parameters from above, a 0.5\degree\ tilt of the linear polarization away from the desired orientation would introduce a $\sim4$ Hz overall shift in the clock transition frequency, and a much smaller differential shift spatially across the lattice owing to variation in the light intensity of the Gaussian-focused beams.  Standard methods for reducing and calibrating this residual shift, including measuring the variation of the clock frequency with lattice-light intensity, should allow the systematic uncertainty to be reduced to an acceptable level~\cite{nicholson_systematic_2015, campbell_fermi-degenerate_2017}.

Additional terms in the light shift, such as the hyperpolarizability and the M1 and E2 polarizabilities would also need to be taken into account, but their effects are small (below 10$^{-18}$ levels in Sr~\cite{porsev_multipolar_2018, ushijima_operational_2018, ton_state-insensitive_2022}), and their consideration is beyond the scope of this paper.

Another significant systematic uncertainty in optical clocks is the blackbody radiation (BBR) shift, which has been the subject of significant past investigation~\cite{porsev_multipolar_2006,safronova_blackbody-radiation_2013}.  We model the BBR shift for the Ti clock line as:
\begin{equation}
    \Delta_{\text{BBR}}= -\kappa \left(\alpha^0_{a\tensor*[^5]{F}{_5}} -\alpha^0_{a\tensor*[^3]{F}{_4}}\right)\left(\frac{T}{300}\right)^4(1+\eta)
\end{equation}
where $\kappa=\frac{1}{2}(831.9[\text{V/m}])^2$ is a constant of proportionality, $\alpha^0_{i}$ is the dc scalar polarizability of the $i$ state of Ti, $T$ is the thermal background temperature measured in K, and $\eta$ is a small dynamical correction omitted in the present work.  The same CI+all-order approach is used to compute dc and dynamic polarizabilities.  In this case, we find that $\alpha^0_{a\tensor*[^5]{F}{_5}}=128.53$ a.u. and $\alpha^0_{a\tensor*[^3]{F}{_4}}=100.39$ a.u., which leads to $\Delta_{\text{BBR}}=-0.24$ Hz at $T=300$ K.  This value is approximately an order of magnitude lower than that in Sr, where the BBR shift is known to be -2.2789 Hz~\cite{lisdat_blackbody_2021}.

The final systematic uncertainty that we consider is the quadratic Zeeman shift (QZS).  For the $^{46,48,50}\text{Ti}$ isotopes, the effect will be small since it will arise only from the mixing of neighboring fine structure states, whereas in atoms with nonzero nuclear spin, a stronger QZS arises from mixing of hyperfine states.  For the states in the Ti clock, the QZS of the $m_J=0$ sublevels are $\Delta_{\text{QZS}}^{(a\tensor*[^3]{F}{_4})}=0.129[\text{Hz/G}^2]B^2$ and $\Delta_{\text{QZS}}^{(a\tensor*[^5]{F}{_5})}=0.434[\text{Hz/G}^2]B^2$, and the QZS on the transition is thus $\Delta_{\text{QZS}}=0.305[\text{Hz/G}^2]B^2$.  Given that a Ti clock must operate at a magnetic field well below $60$ mG to suppress dipolar relaxation, the QZS of the clock transition will be below 1 mHz.  This is approximately an order of magnitude lower than the QZS that is present in Sr optical lattice clocks, of almost 10 mHz~\cite{nicholson_systematic_2015, campbell_fermi-degenerate_2017}.

Altogether, we have shown that laser-cooled Ti is an attractive choice for realizing a telecommunications-band optical atomic clock.  Operating Ti clocks on several of the available telecommunications-band optical transitions would allow for clock comparisons as a powerful method for identifying and reducing systematic corrections.  We have advanced atomic structure calculations to determine critical properties of such clocks, including identifying magic wavelengths for optical trapping, estimating clock transition widths and line strengths, and determining that the BBR shift for Ti clock transitions is an order of magnitude smaller than the shift that dominates current Sr-based clock systematics ~\cite{nicholson_systematic_2015, campbell_fermi-degenerate_2017}.  We also describe potential effects of, and mitigation measures against, magnetic dipole-dipole interactions.  These measures are relevant to other potential applications of dipole-interacting atoms and molecules for precision measurement.

\begin{acknowledgments}

We thank Mikhail Kozlov, Andrey Bondarev, and Ilya Tupitsyn for helpful discussions of polarizability computations.  This work is supported by a collaboration between the US DOE and other Agencies.  This material is based upon work supported by the U.S.\ Department of Energy, Office of Science, National Quantum Information Science Research Centers, Quantum Systems Accelerator.  Additional support is acknowledged from the ONR (Grant Nos.\ N00014-20-1-2513 and N00014-22-1-2280), NSF (PHY-2012068 and the QLCI program through Grant No.\ OMA-2016245), and European Research Council (ERC) under the European Union’s Horizon 2020 research and innovation program (Grant No.\ 856415).
This research was supported in part through the use of University of Delaware HPC Caviness and DARWIN computing systems: DARWIN - A Resource for Computational and Data-intensive Research at the University of Delaware and in the Delaware Region, Rudolf Eigenmann, Benjamin E. Bagozzi, Arthi Jayaraman, William Totten, and Cathy H. Wu, University of Delaware, 2021~\cite{udel}.

\end{acknowledgments}

\bibliography{ti_clock_refs}

\end{document}


\preprint{APS/123-QED}

\title{Supplemental Material for\\ ``Optical Telecommunications-Band Clock based on Neutral Titanium Atoms'' }

\author{Scott Eustice}%
    \affiliation{Department of Physics, University of California, Berkeley, CA 94720}
    \affiliation{Challenge Institute for Quantum Computation, University of California, Berkeley, CA 94720}
\author{Dmytro Filin}
    \affiliation{Department of Physics and Astronomy, University of Delaware, Newark, DE 19716}
\author{Jackson Schrott}
    \affiliation{Department of Physics, University of California, Berkeley, CA 94720}
    \affiliation{Challenge Institute for Quantum Computation, University of California, Berkeley, CA 94720}
\author{Sergey Porsev}
    \affiliation{Department of Physics and Astronomy, University of Delaware, Newark, DE 19716}
\author{Charles Cheung}
    \affiliation{Department of Physics and Astronomy, University of Delaware, Newark, DE 19716}
\author{Diego Novoa}
    \affiliation{Department of Physics, University of California, Berkeley, CA 94720}
    \affiliation{Challenge Institute for Quantum Computation, University of California, Berkeley, CA 94720}
\author{Dan M. Stamper-Kurn}
    \affiliation{Department of Physics, University of California, Berkeley, CA 94720}
    \affiliation{Challenge Institute for Quantum Computation, University of California, Berkeley, CA  94720}
    \affiliation{Materials Science Division, Lawrence Berkeley National Laboratory, Berkeley, CA 94720}
\author{Marianna S. Safronova}
    \affiliation{Department of Physics and Astronomy, University of Delaware, Newark, DE 19716}
    \affiliation{Joint Quantum Institute, National Institute of Standards and Technology and the University of Maryland, College Park, Maryland 20742}
\date{\today}

\maketitle

\section{Calculation of E1, M1, and E2 Transition Rates in Ti I\label{sec:transition_elements}}

\subsection{Theoretical framework: CI+all-order method\label{subsec:theory_background}}
We use the CI+all-order method \cite{safronova_development_2009} that combines linearized coupled cluster and configuration interaction (CI) approaches.  In this method, the electrons are separated into the $1s^22s^22p^63s^23p^6$ core and four remaining valence electrons.  First, the coupled cluster method is used to construct the effective Hamiltonian $H_{\rm{eff}}$ that accounts for core and core-valence correlations and can be constructed using second-order many-body perturbation theory in the CI + MBPT method or the coupled cluster approach (CI + all-order method).  The CI method is used to correlate the remaining four valence electrons using this effective Hamiltonian rather than the usual bare Hamiltonian.  This procedure effectively includes all types of correlation effects in the core and valence spaces.

The CI wave function is constructed as a linear combination of all distinct states of a specified angular momentum $J$ and parity,
\begin{equation}
\psi_J=\sum_{i}c_i \Phi_i ,
\end{equation} 
where $\{\Phi_i\}$ is the set of Slater determinants generated by exciting electrons from the reference configuration to higher orbitals.  The many-electron Schrödinger equation can be written as 
\begin{equation}
H_{\rm{eff}}\Psi=E\Psi,\label{e2}
\end{equation} 
where the effective Hamiltonian has the form 
\begin{equation}
H_{\rm{eff}}=H_{\rm{CI}}+\Sigma.
\end{equation}
Here, $H_{\rm{CI}}$ is the CI Hamiltonian described by the equation 
\begin{equation}
H_{\rm{CI}}=E_{core}+\sum_{i>N_{core}}h_{i,CI}+\sum_{j>i>M_{core}}V_{ij},
\end{equation}
where $E_{core}$ is the energy of the frozen core, $N_{core}$ is the number of core electrons, $h_{i,CI}$ represents the kinetic energy of the valence electrons and their interaction with the central field and $V_{ij}$ accounts for the valence–valence correlations.

The core-valence correlation potential, \[ \Sigma=\Sigma_1+\Sigma_2, \]
is obtained from the all-order method.  Here, $\Sigma_1$ and $\Sigma_2$ are the one- and two-electron parts of the core–valence correlation potential, respectively.  After Eq.~(\ref{e2}) is solved using the CI technique and the wave functions are obtained, they are used to calculate matrix elements of the electric-dipole, magnetic-dipole, electric-quadrupole, and other one-electron operators.

\subsection{Energy level calculation\label{subsec:energy_levels}}
When applying the CI+all-order method to atomic Ti, we used a a $V^{N-4}$ potential of the $1s^22s^22p^63s^23p^6$ frozen core.  We solve Dirac-Hartree-Fock equations in this potential to generate $3d$, $4s$, $4p$, $5s$, $4d$, $5p$, and $4f$ orbitals.  All other orbitals are constructed in a spherical cavity of 40 a.u.\ using B-splines.

The set of CI configurations has to be constructed separately for even and odd states.  We carry out several calculations with increasing number of configurations to ensure convergence of the CI with the number of included configurations.  For even states, we find it sufficient to make all possible single and double excitations to a $20spd18f16g$ basis starting from $4s^23d^2$, $4s3d^25s$, $4s^23d4d$, $3d^24p^2$, $3d^25s^2$, $4s3d^3$, $4s3d^24d$, $3d^35s$, and $3d^25s4d$ configurations.  We verified that a subset of triple excitations give a negligible contribution.  For odd states, the CI configuration space is sufficiently saturated by the single and double excitations to the same large basis from the $4s3d^24p$, $3d^34p$, $4s3d^25p$, $3d^24p5s$, $4s3d4p4d$, $3d^25s5p$, and $3d^24p4d$ configurations.

We have carried out calculation of energies for 85 levels and compared them with experiment.  Most of the theoretical energies differ from experimental values by only 0.1-2.5\%.  We present results only for the lines and levels that can be used to pump titanium optically into the metastable state to perform laser cooling and to drive the relevant clock transitions.  The selected energy levels in cm$^{-1}$ are listed in Tables~\ref{tab:Eeven} and~\ref{tab:Eodd}.

\begin{table}[t]
\caption{Comparison of theoretical even energy levels (in cm$^{-1}$) with experiment \cite{kramida_nist_1999}.}
\label{tab:Eeven}
\setlength{\extrarowheight}{3.5pt}
  \begin{ruledtabular}
  \begin{tabular}{lcrrrr}
  
Configuration & Term & Expt& Theory & Diff &Diff \% \\
\hline
$3d^24s^2$ & $a\tensor*[^3]{F}{_2}$ & 0 & 0 & 0 & \\ 
& $a\tensor*[^3]{F}{_3}$ & 170 & 177 & 7 & 4.1\% \\
& $a\tensor*[^3]{F}{_4}$ & 387 & 396 & 9 & 2.3\% \\
$3d^3(^4F)4s$ & $a\tensor*[^5]{F}{_1}$ & 6557 & 6374 & -183 & -2.8\% \\	
& $a\tensor*[^5]{F}{_2}$ & 6599 & 6416 & -183 & -2.8\% \\
& $a\tensor*[^5]{F}{_3}$ & 6661 & 6477 & -184 & -2.8\% \\
& $a\tensor*[^5]{F}{_4}$ & 6743 & 6557 & -186 & -2.8\% \\
& $a\tensor*[^5]{F}{_5}$ & 6843 & 6652 & -191 & -2.8\% \\
$3d^3(^2G)4s$ & $a\tensor*[^3]{G}{_5}$ & 15220 & 15497 & 276 & 1.8\% \\
$3d^3(^2H)4s$ & $a\tensor*[^3]{H}{_5}$ & 18141 & 18450 & 308 & 1.7\% \\
& $a\tensor*[^3]{H}{_6}$ & 18193 & 18498 & 306 & 1.7\% \\
$3d^3(^2H)4s$ & $a\tensor*[^1]{H}{_5}$ & 20796 & 21171 & 375 & 1.8\% \\

\end{tabular}
\end{ruledtabular}

\end{table}

\begin{table}[t]
\caption{ Comparison of theoretical odd energy levels (in cm$^{-1}$) with experiment \cite{kramida_nist_1999}.}
\label{tab:Eodd}
\setlength{\extrarowheight}{3.5pt}
  \begin{ruledtabular}
  \begin{tabular}{lcrrrr}
  
Configuration & Term & Expt& Theory & Diff &Diff \% \\
\hline
$3d^2(^3F)4s4p(^3P^o)$ & $z\tensor*[^5]{G}{^o_6}$ & 16459 &16454& -5& 0.0\%\\
$3d^2(^3F)4s4p(^3P^o)$ & $z\tensor*[^1]{S}{^o_0}$ & & 24174 &   &   \\   
$3d^2(^3P)4s4p(^3P^o)$ & $\tensor*[^5]{D}{^o_4}$ & 25927 &26081& 154&0.6\% \\   
$3d^3(^4F)4p$ & $y\tensor*[^5]{G}{^o_6}$ & 26911 &26982& 71& 0.3\% \\   

\end{tabular} 
\end{ruledtabular}
\end{table}

We compute the expectation values $\langle L^2\rangle$ and $\langle S^2\rangle$, where $L$ and $S$ are the total electron orbital and spin angular momentum operators, to obtain approximate quantum numbers $L$ and $S$, where $\langle L^2\rangle = L(L+1)$ and $\langle S^2\rangle = S(S + 1)$, which allowed us to unambiguously identify all terms in Tables~\ref{tab:Eeven} and \ref{tab:Eodd}.  As a result, we identify a level, 3d$^2$4s4p	$^1$S$^o_0$, not listed in the NIST database.  This level is included in Table~\ref{tab:Eodd}.

\subsection{Optical clock transitions\label{subsec:opt_clock_transitions}}

\begin{table*}[h]
\caption{Results of calculations for the $a\tensor*[^3]{F}{_4}\to a\tensor*[^5]{F}{_5}$ clock transitions of Ti.  Wavelength, $\lambda$, (in units of nm), E2 and M1 reduced matrix elements (ME) (in a.u. for E2 transitions, in $\mu_B$ for M1 transitions), transition rates (in $\times 10^{-7}$s$^{-1}$), and branch ratio.}
\label{tab:clock_trans} 
\setlength{\extrarowheight}{5pt}
  \begin{ruledtabular}
  \begin{tabular}{lllllllllll}

Upper & Term &Level& Lower & Term & Level& $\lambda$& & ME& Tr. rate & Branching ratio\\
\hline
$3d^3(^4F)4s$ & $a\tensor*[^5]{F}{_5}$ & 6843 & $3d^24s^2$ & $a\tensor*[^3]{F}{_3}$ & 170 & 1498.6 & E2 & 0.0472(7) & 3.00(9) & 0.00838(25) \\
&&&& $a\tensor*[^3]{F}{_4}$ & 387 & 1548.9 & M1 & 0.0010(5) & 7(6) & 0.020(17) \\
&&&&&&& E2 & 0.140(4) & 22.4(1.3) & 0.063(4) \\
&&& $3d^3(4F)4s$ & $a\tensor*[^5]{F}{_4}$ & 6743 & 99794.4 & M1	& 3.632(10) & 325.43(18) & 0.909(16)\\
&&&&&&& E2 & 1.748(18) & $3.14(8)\times 10^{-6}$ & $8.78(22)\times 10^{-9}$\\
$3d^24s^2$ & $a\tensor*[^3]{F}{_4}$ & 387 & $3d^24s^2$ & $a\tensor*[^3]{F}{_3}$ & 170 & 46138. & M1	& 2.597(0.026) & 2059(53) & 1
\end{tabular}
\end{ruledtabular}

\end{table*}

\begin{table*}[h]

\caption{Wavelengths (in nm), electric-dipole reduced matrix elements D$_{tot}$ (in a.u.), transition rates (in s$^{-1}$), and branching ratios for cooling transitions of Ti I.  The D$_{tot}$ values (in a.u.) are calculated with CI+all-order method and include the random-phase approximation (RPA), the core-Brueckner ($\sigma$), structural radiation (SR), two-particle (2P), and normalization (Norm) corrections.  For approximate values (indicated by a $\sim$ symbol), the precise value of the matrix element is highly uncertain, and the reported value should be interpreted as correct only within an order of magnitude.}

\label{tab:laser_cool}
\setlength{\extrarowheight}{5pt}
  \begin{ruledtabular}
  \begin{tabular}{lllllllllll}

Upper & Term &Level& Lower & Term & Level& $\lambda$& & D$_{tot}$& Tr. rate & Branch ratio\\
\hline
$3d^3(^4F)4p$ &	$y\tensor*[^5]{G}{^o_6}$ &	26911 &	$3d^3(^4F)4s$ &	$a\tensor*[^5]{F}{_5}$	&	6843	& 498 &	E1	&	7.337(12) & 6.780(23)$\times 10^7$ & 0.9999989(5)	\\
&&& $3d^3(^2G)4s$ &	$a\tensor*[^3]{G}{_5}$ &15220& 855&E1&$\sim0.007$& $\sim11$& $\sim2\times 10^{-7}$
\\
&&&	$3d^3(^2H)4s$ & $a\tensor*[^3]{H}{_5}$ &	18141	& 1140 &	E1&0.0055(14) &3.2(1.4)&4.7(2.0)$\times 10^{-8}$
	\\
&&&$3d^3(^2H)4s$ & $a\tensor*[^3]{H}{_6}$&18193& 1147&E1	&0.025(4) &62(19) &9(3)$\times 10^{-7}$	\\
&&&$3d^3(^2H)4s$ & $a\tensor*[^1]{H}{_5}$ &20796& 1635&E1&$\sim0.0004$& $\sim 0.005$& $\sim 7\times 10^{-11}$	\\
$3d^2(^3F)4s4p(^3P^o)$ & $z\tensor*[^5]{G}{^o_6}$ &	16459 &	$3d^3(^4F)4s$	&	$a\tensor*[^5]{F}{_5}$	&	6843	& 1040 &	E1	&	0.86(3) & 1.03(8)$\times 10^5$ & 0.99999969(2)	\\
&&&	$3d^3(^2G)4s$ & $a\tensor*[^3]{G}{_5}$ &15220& 8076&E1&	0.01040(10)&0.0318(25)&3.1(2)$\times 10^{-7}$	\\

\end{tabular}
\end{ruledtabular}

\end{table*}

\begin{table*}[h]
\caption{Wavelengths (in nm), calculated and observed transition rates (in s$^{-1\times10^6}$), and branching ratios for optical pumping of Ti.  The upper state for all transitions is the $3d^2(^3P)4s4p(^3P^o)$ $\tensor*[^5]{D}{^o_4}$ level, at 25967 cm$^{-1}$.}
\setlength{\extrarowheight}{5pt}
\begin{ruledtabular}
\begin{tabular}{lllllll}
    Lower & Term & Level & $\lambda$ & Tr. rate, theory & Tr. rate, lit & Branch ratio \\ \hline
    $3d^24s^2$ & $a\tensor*[^3]{F}{_3}$ & 170 & 388.2 & 0.13(6) & 0.28(6) & 0.049 \\
    & $a\tensor*[^3]{F}{_4}$ & 387 & 391.5 & 1.8(8) & 2.11(28) & 0.68\\
    $3d^3(^4F)4s$ & $a\tensor*[^5]{F}{_3}$ & 6661 & 519.1 & 0.00036(12) &  & 0.00014 \\
    & $a\tensor*[^5]{F}{_4}$ & 6743 & 521.3 & 0.08(8) & 0.31(16) & 0.030 \\
    & $a\tensor*[^5]{F}{_5}$ & 6843 & 524.0 & 0.4(4) &  & 0.17 \\
    & $b\tensor*[^3]{F}{_4}$ & 11777 & 706.7 & 0.06(3) &  & 0.02 \\
    $3d^3(^4P)4s$ & $a\tensor*[^5]{P}{_3}$ & 14106 & 845.9 & 0.129(4) &  & 0.049 
\end{tabular}
\end{ruledtabular}
\end{table*}

 We study forbidden transitions between the $a\tensor*[^3]{F}{}$ and $a\tensor*[^5]{F}{}$ terms to identify the most suitable clock transition.  The main text summarizes the properties of the clock transitions.  The total clock transition linewidth accounts not only for spontaneous decay on the clock transition itself, but also for decay of the upper and lower levels of the clock transition to other states, leading to an overall clock transition linewidth that is larger than the spontaneous decay rate on the clock transition alone.  In Table~\ref{tab:clock_trans}, we list the contributions that determine the linewidth of the $a^3F_4\to a^5F_5$ clock transition.  We note that the clock transition linewidth is dominated by M1 decays of both the lower and upper states to other fine structure states within their respective manifolds.  The same calculation was performed to calculate the linewidth of all other clock transitions. 
 The random-phase approximation (RPA) corrections are included to the effective electric quadupole and magnetic dipole operators, see, for example Ref.\ \cite{safronova_blackbody-radiation_2013}.  Such effective operators account for the core-valence correlations in analogy with the effective Hamiltonian $H_{\rm{eff}}$ discussed above.  The uncertainties in the values of matrix elements were estimated as difference between values obtained using CI+all-order and CI+MBPT methods.

\subsection{Laser cooling transitions\label{subsec:laser_cooling}}

Ref.~\cite{eustice_laser_2020} identified two candidate transitions on which Ti may be laser cooled.  To support ongoing experimental efforts to realize laser cooling of Ti, we characterized these two transitions theoretically using our atomic-structure calculations described herein.  Specifically, we calculated the strengths of the two electric-dipole laser cooling transitions, and also, critically to experimental efforts, determined the small leakage rate (branching ratio) out of the laser cooling transitions.  Results for two of the cooling transitions, $3d^3(\tensor*[^4]{F}{})4p$ $y\tensor*[^5]{G}{^{^\circ}_6}$ - $3d^3(\tensor*[^4]{F}{})4s$ $a\tensor*[^5]{F}{_5}$ at $\lambda=498$ nm and $3d^3(\tensor*[^3]{F}{})4s4p(\tensor*[^3]{P}{^o})$ $z\tensor*[^5]{G}{^o_6}$ - $3d^3(\tensor*[^4]{F}{})4s$ $a\tensor*[^5]{F}{_5}$ at $\lambda=1040$ nm are listed in Table~\ref{tab:laser_cool}).

In Ref.~\cite{eustice_laser_2020} it was noted that while there do exist other even parity states to which a Ti atom in the excited state of the laser cooling transition can decay, such scattering would be strongly suppressed because the transitions are spin forbidden.  Our calculations confirm this expectation.  Indeed, we find the branching to those states is exceptionally low, at the $10^{-6}$ level for the 498 nm transition and at the $3\cdot10^{-7}$ level for the 1040 nm transition.  These are low enough branching ratios to enable the typical tools of modern ultracold atomic physics, including both Doppler and sub-Doppler cooling techniques and single-atom fluorescenence detection in quantum-gas microscopes or optical tweezers, without the need for additional repumping lasers.

To obtain transition rates and branching ratios for the cooling transitions (see Table \ref{tab:laser_cool}) we use the electric-dipole reduced matrix elements calculated with the effective electric-dipole operator in the random-phase approximation.  We also considered other correction to the E1 operator beyond RPA: the core-Brueckner ($\sigma$), structural radiation (SR), two-particle (2P), and normalization (Norm) corrections \cite{dzuba_using_1998,porsev_electric-dipole_1999,porsev_calculation_1999}.  As has been noted for the case of Sr~\cite{safronova_blackbody-radiation_2013}, these corrections cannot be omitted at the 1\% level of accuracy.

In Table \ref{tab:laser_cool} we include transition matrix elements, $D_{tot}$, obtained taking into account all the corrections mentioned above.  The uncertainties were estimated by taking the difference between the values obtained using CI+all-order and CI+MBPT methods.  For very small matrix elements (marked by the tilde symbol), we provide approximate values without explicit evaluation of errors. For these, the error should be assumed to be at the same order of magnitude as the quantity itself.

\subsection{Optical pumping transitions\label{subsec:optical_pumping}}

Gas phase Ti atoms may be produced with a thermal source that operates between 1200\degree C to 1800\degree C. 
 At these temperatures, a low population of the Ti atoms produced would be in the $a\tensor*[^5]{F}{_5}$ laser-coolable state.  It would therefore be necessary to transfer atoms to this state from the $a\tensor*[^3]{F}{}$ ground state manifold via optical pumping.  Through our calculations, we examined multiple potential excited states that could be used to transfer population efficiently to the $a\tensor*[^5]{F}{_5}$ state. We found the $3d^2(\tensor*[^3]{P}{})4s4p(\tensor*[^3]{P}{^o})$ $y\tensor*[^5]{D}{^o_4}$ state has a significant branching ratio to both the $a\tensor*[^5]{F}{_5}$ state and $a\tensor*[^3]{F}{}$ states, allowing for optical pumping of atoms from the ground states to the laser cooling state. The transition wavelengths are also amenable to current laser technology.

\section{Optical Clock Rabi Frequencies}

The optical clock transitions of Ti can be driven by either the M1 or E2 multipoles of the electromagnetic field, both of which must be accounted for when calculating the overall Rabi frequency.  In each of the transitions we highlighted in the main text ($m_J=0\to m'_J=0$, $J=4\to J'=4,5$) only one of the two multipoles is allowed.

Given a reduced M1 matrix element, $D_{M1}$, for a transitions between lower and upper levels $J$ and $J'$, and a magnetic field $\mathbf{B}$ of the clock laser, the M1 Rabi frequency between a lower and upper sublevel $m_J$ and $m'_J$ is given by
\begin{multline}
    \Omega_{M1} = -\frac{D_{M1}}{\hbar}(-1)^{J'-m'_J}%
    \hat{\mathbf{e}}^*_{m'_J-m_J}\cdot\mathbf{B}\\%
    \times\begin{pmatrix}%
        J & 1 & J'\\
        m_J & m'_J-m_J & -m'_J
    \end{pmatrix} .%
\end{multline}

For an electric quadrupole transition with reduced matrix element $D_{E2}$ driven by a plane-wave clock laser field $\mathbf{E}$ with polarization $\boldsymbol{\varepsilon}$ and propagation wavevector $\mathbf{k}$, the Rabi frequency is
\begin{multline}
    \Omega_{E2} = \frac{iD_{E2}}{2\hbar}(-1)^{J'-m'_J}%
    \begin{pmatrix}%
        J & 2 & J'\\
        m_J & m'_J-m_J & -m'_J
    \end{pmatrix}\\%
    \times\sum_{i,j}M_{ij}(m'_J-m_J)k_i\varepsilon_j.
\end{multline}
Here, $M_{ij}(q)$ is a geometric factor given by
\begin{equation}
    M_{ij}(q) = (-1)^q\sqrt{5}\sum_{q_1,q_2}(\mathbf{\hat{e}}_i\cdot\mathbf{\hat{e}}^*_{q_1})(\mathbf{\hat{e}}_j\cdot\mathbf{\hat{e}}^*_{q_1})%
    \begin{pmatrix}%
        1 & 1 & 2\\
        q_1 & q_2 & -q
    \end{pmatrix} ,%
\end{equation}
where $\mathbf{\hat{e}}_i$ and $\mathbf{\hat{e}}_q$ are the cartesian and spherical basis vectors respectively.

\section{Dynamical Polarizability of Ti I\label{sec:dyn_polariz}}

\begin{table*} [h]
\caption{Contributions to the static electric-dipole polarizability $\alpha_0$ with the appropriate reduced matrix elements $D$ (in a.u.) of the $3d^24s^2$ $a\tensor*[^5]{F}{_5}$ level.  For comparison, the experimental and theoretical energy levels (in cm$^{-1}$) are shown.}
\label{tab:5F5}
\setlength{\extrarowheight}{5pt}
  \begin{ruledtabular}
  \begin{tabular}{llccccc}
Configuration & Term & Exp. Level&Theor. Level&Diff \% & D &Contrib. to $\alpha_0$ \\
\hline
$3d^2(^3F)4s4p(^3P^o)$ &$z\tensor*[^5]{G}{^o_6}$& 16459&16454& 0.0\% &0.86(3) &1.00(8)\\
&$z\tensor*[^5]{F}{^o_5}$& 17215&17185& -0.2\% &2.48(6) &7.7(4)\\
 &$z\tensor*[^5]{D}{^o_4}$& 18695&18769& 0.4\% &2.08(3) &4.77(16) \\
$3d^3(^4F)4p$  &$y\tensor*[^5]{G}{^o_5}$& 26773&26850& 0.3\%&2.299(5) &3.48(2)\\
 &$y\tensor*[^5]{G}{^o_6}$& 26911&26982& 0.3\%&7.837(12) &40.16(12)\\
 &$y\tensor*[^5]{F}{^o_4}$& 28788&29061& 1.0\%&2.410(7) &3.45(2)\\
 &$y\tensor*[^5]{F}{^o_5}$& 28896&29164& 0.9\%&7.251(6) &31.07(5)\\ 
 &$x\tensor*[^5]{D}{^o_4}$& 30060&30474& 1.4\%&6.723(12) &25.24(12)\\ 
\hline
Other & &   &   & & & 11.7   \\
Total & &   &   & & & 128.5(2.0) \\
\end{tabular}
\end{ruledtabular}
\end{table*}  

\begin{table*} [h]
\caption{Contributions to the static electric-dipole polarizability $\alpha_0$ with the appropriate reduced matrix elements $D$ (in a.u.) of the $3d^24s^2$ $a\tensor*[^3]{F}{_4}$ level.  For comparison, the experimental and theoretical energy levels (in cm$^{-1}$) are shown.}
\label{tab:3F4}
\setlength{\extrarowheight}{5pt}
  \begin{ruledtabular}
  \begin{tabular}{llccccc} 
Configuration     & Term & Exp. Level&Theor. Level&Diff \% &  D &Contrib. to $\alpha_0$  \\
\hline
$3d^2(^3F)4s4p(^3P^o)$  &$z\tensor*[^3]{F}{^o_4}$& 19574&19632&  0.3\% &1.682(13) &2.39(4)\\
 &$z\tensor*[^3]{D}{^o_3}$& 20126&20115&  -0.1\% &1.467(9) &1.774(22)\\
 &$z\tensor*[^3]{G}{^o_5}$& 21740&22025&  1.3\% &1.297(4) &1.263(8) \\
 &$z\tensor*[^1]{G}{^o_4}$& 24695&25163&  1.9\% &1.1(4) &0.8(6) \\
$3d^2(^3F)4s4p(^1P^o)$  &$y\tensor*[^3]{F}{^o_3}$& 25227&25351&  0.5\%&1.479(13)  &1.423(25)\\
 &$y\tensor*[^3]{F}{^o_4}$& 25388&25531&  0.6\%&3.9(3)  &9.8(1.8)\\
$3d^3(^4F)4p$   &$y\tensor*[^3]{D}{^o_3}$& 25644&25591&  -0.2\%&3.12(17)  &6.3(7)\\ 
$3d^2(^1D)4s4p(^3P^o)$  &$x\tensor*[^3]{F}{^o_4}$& 27026&27325&  1.1\%&3.36(11)  &6.8(5)\\ 
$3d^2(^3F)4s4p(^1P^o)$  &$y\tensor*[^3]{G}{^o_5}$& 27750&27846&  0.35\%&4.65(25)  &12.7(1.4)\\ 
 &$w\tensor*[^3]{D}{^o_3}$& 29912&30145&  0.8\%&2.983(17)  &4.86(5)\\
$3d^2(^1G)4s4p(^3P^o)$  &$x\tensor*[^3]{G}{^o_5}$& 30039&30349&  1.0\%&4.416(10)  &10.53(5)\\ 
$3d^3(^4F)4p$   &$w\tensor*[^3]{G}{^o_5}$& 31629&32053&  1.3\%&4.61(20)  &10.9(1.0)\\
$3d^2(^1G)4s4p(^3P^o)$  &$v\tensor*[^3]{F}{^o_4}$& 34205&34742&  1.6\%&3.60(14) &6.1(5)\\ 
$3d^3(^2D_2)4p$ &$u\tensor*[^3]{F}{^o_3}$& 37744&38871& 3.0\%&2.74(13) &3.2(3)\\
$3d^2(^3P)4s4p(^1P^o)$ &$u\tensor*[^3]{D}{^o_3}$& 38159&38909& 2.0\%&1.74(5) &1.28(7)\\
$3d^3(^2G)4p$  &$t\tensor*[^3]{F}{^o_4}$& 38671&39560& 2.3\%&2.6(4) &2.9(8)\\
$3d^3(^2D_2)4p$ &$s\tensor*[^3]{D}{^o_3}$& 39715&40515& 2.0\%&1.5(4) &0.9(5)\\
$3d^24s(^4F)5p$ &$o\tensor*[^3]{D}{^o_3}$& 44234&45270& 2.3\%&2.11(8) &1.61(12)\\
\hline
Other & &   &   & & & 14.9\\
Total & &   &   & & & 100.4(1.8) \\
\end{tabular}
\end{ruledtabular}
\end{table*}   

In order to find the magic wavelengths of the clock transitions, it is necessary to calculate the polarizability of Ti in both the $a\tensor*[^3]{F}{_4}$ and $a\tensor*[^5]{F}{_5}$ states over a wide range of frequencies.  Performing this calculation directly remains a computational challenge for complex atoms like Ti.  Direct calculation requires the inversion of huge matrices --- 350,000 $\times$ 350,000 in our case --- which is computationally intractable.  Instead, we use an iterative approach included in the pCI Code Package~\cite{cheung_scalable_2021} implementing the CI+all-order technique to calculate the polarizability of the considered states of Ti.  This approach allows us to get accurate results using the inversion of smaller matrices (15,000 $\times$ 15,000), making the task feasible.  Unfortunately, this method does not work at all frequencies, owing to the potential divergence of the iterative process.  We successfully obtained static polarizabilities for both $3d^24s^2$ $a\tensor*[^3]{F}{_4}$ and $3d^3(^4F)4s$ $a\tensor*[^5]{F}{_5}$ levels, but for dynamic polarizabilities the calculation diverged for wavelengths shorter than 750 nm (900 nm) for the $a\tensor*[^3]{F}{_4}$ ($a\tensor*[^5]{F}{_5}$) level.  To overcome this, we used a combination of the CI+all-order technique and the sum-over-states method to calculate the scalar, vector and tensor polarizability from 1100 nm to 400 nm.

The sum-over-states method involves using only bound states of an atom, and there is always some inaccuracy due to missing contributions from continuum states and bound states not included in the calculation.  For the best accuracy, it is advisable to use as many states as possible in the sum.  However, there are always limits on the accuracy of calculations of highly excited states.  To balance these trade-offs, we use in the sum-over-state method for the lower lying states that contribute the majority of the dc polarizability but still can be properly calculated.  For this purpose, we used 73 states to obtain the polarizability of the $3d^24s^2$ $a\tensor*[^3]{F}{_4}$ level and 49 states for the polarizability of the $3d^3(^4F)4s$ $a\tensor*[^5]{F}{_5}$ level.  The most important contributions to both polarizabilities are shown in Tables~\ref{tab:5F5}, \ref{tab:3F4}.  The polarizability is generally divided into three terms: $\alpha^S$ - scalar, $\alpha^V$-vector, and $\alpha^T$ - tensor polarizabilities.  They are represented for an arbitrary state $i$ as follows~\cite{le_kien_dynamical_2013,bai_angle-dependent_2022}:
\begin{equation} \label{eq:pol_scalar}
   \alpha^S_i(\omega) = \frac{2}{3(2J_i+1)}\sum_n \frac{(E_n-E_i)|\langle n ||D||i\rangle|^2 }{(E_n-E_i)^2-\omega^2}
\end{equation}
\begin{equation}
    \begin{split}
        \alpha^V_i(\omega) =  C_1\sum_n (-1)^{J_n+J_i}\begin{Bmatrix}
        1   & 1   & 1 \\
        J_i & J_i & J_n
        \end{Bmatrix}\\ \frac{\omega |\langle n ||D||i\rangle |^2}{(E_n-E_i)^2-\omega^2}
    \end{split}   
\end{equation}
\begin{equation}
    \begin{split}
       \alpha^T_i(\omega) =  C_2\sum_n (-1)^{J_n+J_i}\begin{Bmatrix}
       1   & 1   & 2 \\
       J_i & J_i & J_n
       \end{Bmatrix}\\ \frac{(E_n-E_i)|\langle n ||D||i\rangle |^2}{(E_n-E_i)^2-\omega^2}, 
    \end{split}
\end{equation}
where
\[ C_1=-2 \sqrt{\frac{6J_i}{(J_i+1)(2J_i+1)}} \]
and
\[ C_2=4 \sqrt{\frac{5J_i(2J_i-1)}{6(J_i+1)(2J_i+1)(2J_i+3)}}. \]
The index $n$ refers to the states in the sum-over-states that contribute to the polarizability of the state $i$; $E_{n,i}$ is the energy and $J_{n,i}$ is the total angular momentum of the state; $\langle n||D||i\rangle$ is the reduced matrix element between the two states, and $\omega$ is the frequency of the external electric field.  The total dynamic polarizability can be expressed as follows:
\begin{equation} \label{eq:pol_dyn_full}
   \begin{split}
   \alpha_i(\omega) &= \alpha^S_i(\omega) + \varepsilon\cos(\theta_k)\frac{m_{J_i}}{2J_i}\alpha^V_i(\omega) + \\
                    &+\left( \frac{3\cos^2\theta_p-1}{2}\right)\frac{3m^2_{J_i}-J_i(J_i+1)}{J_i(2J_i-1)}\alpha^T_i(\omega),
   \end{split}
\end{equation}
where $\varepsilon$ is the ellipticity of the polarization, $\theta_k$ is the angle between the direction of propagation of the light and the quantization axis, $\theta_p$ is the angle between the polarization of the light and the quantization axis, and $m_{J_i}$ is a magnetic quantum number.

To obtain an accurate value of dynamic polarizability with the sum-over-states method, one has to estimate the residual contribution to the polarizability from states that are not included in summation.  We made this estimate using the fact that $|\omega/(E_n -E_i)| \ll 1$ for all such residual states in the range of wavelengths that we are considering.  Indeed, using this ratio as a small parameter and expanding to the lowest non-zero order in, Eq.~\ref{eq:pol_scalar} can be simplified to
\begin{equation} \label{eq:pol_scal_simpl}
   \alpha^S_i(\omega) = \alpha^{S_N}_i(\omega) + \alpha^{S_{res}}_i(\omega)
\end{equation}
where
\begin{equation} \label{eq:pol_scal_N}
   \alpha^{S_N}_i(\omega) = \frac{2}{3(2J_i+1)}\sum^N_{n=1} \frac{(E_n-E_i)|\langle n |D|i\rangle|^2 }{(E_n-E_i)^2-\omega^2}
\end{equation}
and where $N$ is the number of states used in direct summation. Moreover, we can write 
\begin{equation} \label{eq:pol_s_res}
\alpha^{S_{res}}_i(\omega)= A_0+B_0\, \omega^2
\end{equation}
where $A_0$ and $B_0$ are expansion constants for the long wavelength scalar polarizability.  The same simplification works for $\alpha^V_i(\omega)$ and $\alpha^T_i(\omega)$, by writing $\alpha^{V_N}_i(\omega)$ and $\alpha^{T_N}_i(\omega)$ as a summation as in \ref{eq:pol_scal_N}. This yields the expansion of the long wavelength residuals:
\begin{equation} \label{eq:pol_vt_res}
\begin{split}
\alpha^{V_{res}}_i(\omega)&= B_1\omega \\
\alpha^{T_{res}}_i(\omega)&= A_2+B_2\, \omega^2  
\end{split}
\end{equation}
where $B_1$, $A_2$, and $B_2$ are additional expansion constants.

Using the results of the sum-over-states method, we identified the four candidate magic wavelengths discussed in the main text.  The atoms were assumed to have magnetic quantum numbers $m_J=0$ and the optical field parameters were $\varepsilon=0$ and $\theta_p=\theta_k=90\degree$.  However, the more accurate iterative approach used by the pCI code package was found to not converge at all of the candidate wavelengths.  Using the fact that convergence in the iterative polarizability can be achieved at $\omega=0.043989$ a.u. ($\lambda=1035.8$ nm) and for all smaller values of $\omega$ down to $\omega=0$, we  subtracted the values of $\alpha_i(0.043989)$ and $\alpha_i(0)$ obtained from the sum-over-states method from the corresponding results computed with the pCI code. In this way, we found the residuals $\alpha^{S_{res}}_i(0)$, $\alpha^{V_{res}}_i(0)$, $\alpha^{T_{res}}_i(0)$ and $\alpha^{S_{res}}_i(0.043989\,\mathrm{ a.u.})$, $\alpha^{V_{res}}_i(0.043989\,\mathrm{ a.u.})$, $\alpha^{T_{res}}_i(0.043989\,\mathrm{ a.u.})$. This allowed us to determine the parameters $A_{0,2}$ and $B_{0,1,2}$ in Eqs.~\ref{eq:pol_s_res}, \ref{eq:pol_vt_res} and thus extend the accuracy of the direct iterative polarizability calculation to frequencies where the iterative calculation fails to converge.

The uncertainty on the polarizabilities were obtained by comparing the calculations of the more accurate CI+all-order technique with CI+MBPT at the points where the direct calculation converged.  The difference between the values of the appropriate reduced matrix elements $|\langle n ||D||i\rangle|_{CI+all-order}$ and $|\langle n ||D||i\rangle|_{CI+MBPT}$ is an additional possible inaccuracy of the method and thus is used to estimate the uncertainties of the polarizabilities.  Furthermore, possible errors in the parameters $A_{0,2}$ and $B_{0,1,2}$ were included in the final uncertainties as the difference between the correspondent parameters calculated with the CI+all-order and CI+MBPT methods.

\section{Dipole-Dipole Interactions\label{sec:dip_dip_int}}

As discussed in the main text, dipole-dipole interactions underlie several effects that impact the operation of a Ti atomic clock.  One can distinguish between interactions that are inelastic or elastic in the motional degrees of freedom.  Inelastic interactions can be described as spin relaxation, a dipole-mediated conversion of internal Zeeman energy into external motional energy.  For example, in an optical lattice spin relaxation may couple atoms from lower to higher bands of the lattice.  As discussed in the main text, inelastic motional interactions can be suppressed by trapping atoms in a lattice with a sufficiently large band gap~\cite{de_paz_resonant_2013}.

One may then focus on elastic motional interactions.  We describe these interactions purely in the spin sector using the secular Hamiltonian, which accounts for cycle-averaging over the Larmor precession of the atomic spins~\cite{chomaz_dipolar_2022}.  This secular Hamiltonian, whose form is presented also in the main text, is given for two atoms as
\begin{multline}
    H_{dd}=\frac{\mu_0\mu_B^2}{4\pi}\frac{g_{J_1}g_{J_2}}{r_{12}^3}\left(1-3\cos^2\theta_{12}\right)\\
    \times\left(J^z_1J^z_2-\frac{1}{4}\left(J^+_1J^-_2 + J^-_1J^+_2\right)\right)
\end{multline}
with $1$ and $2$ labelling the two atoms, and $r_{12}$ and $\theta_{12}$ describing their position difference vector and the angle it makes with the quantization axis.

One can consider separately two effects of this secular Hamiltonian.  The first is a spin-elastic interaction, which is diagonal in the separable basis of magnetic sublevel states.  This shift is zero on the $m_J=0$ magnetic sublevels and therefore does not affect a clock based on an $m_J=m'_J=0$ transition.

The second effect is the spin mixing interaction, which is off-diagonal in the separable magnetic sublevel basis.  This interaction conserves the total magnetic quantum number of the two atoms, $m_J^{(\mathrm{tot})}=m_J^{(1)}+m_J^{(2)}$, but changes the magnetic quantum numbers of the individual atoms. In the absence of dipolar interactions, the entire manifold of two-atom states with identical $m_J^{(\mathrm{tot})}$ is degenerate for atoms with equal Zeeman splittings.  Dipolar interactions can then generate significant mixing and energy shifts within this manifold, leading to imprecise measurement of the clock transition.

While several methods have been developed in NMR to control dipolar spin mixing, e.g.\ multiple pulse sequences \cite{choi_robust_2020} and magic angle spinning \cite{polenova_magic_2015}, these techniques are not needed to suppress spin mixing in a Ti clock system.  As described in the main text, spin mixing within the clock states is mitigated by the tensor light shift imposed by the optical lattice beams. From Eq.~\ref{eq:pol_dyn_full}, the tensor light shift includes a part proportional to $m^2_J$, which splits the degeneracy of the aforementioned states and energetically suppresses the spin mixing process.

\begin{figure}[t]
    \centering
    \includegraphics[width=1\linewidth, keepaspectratio]{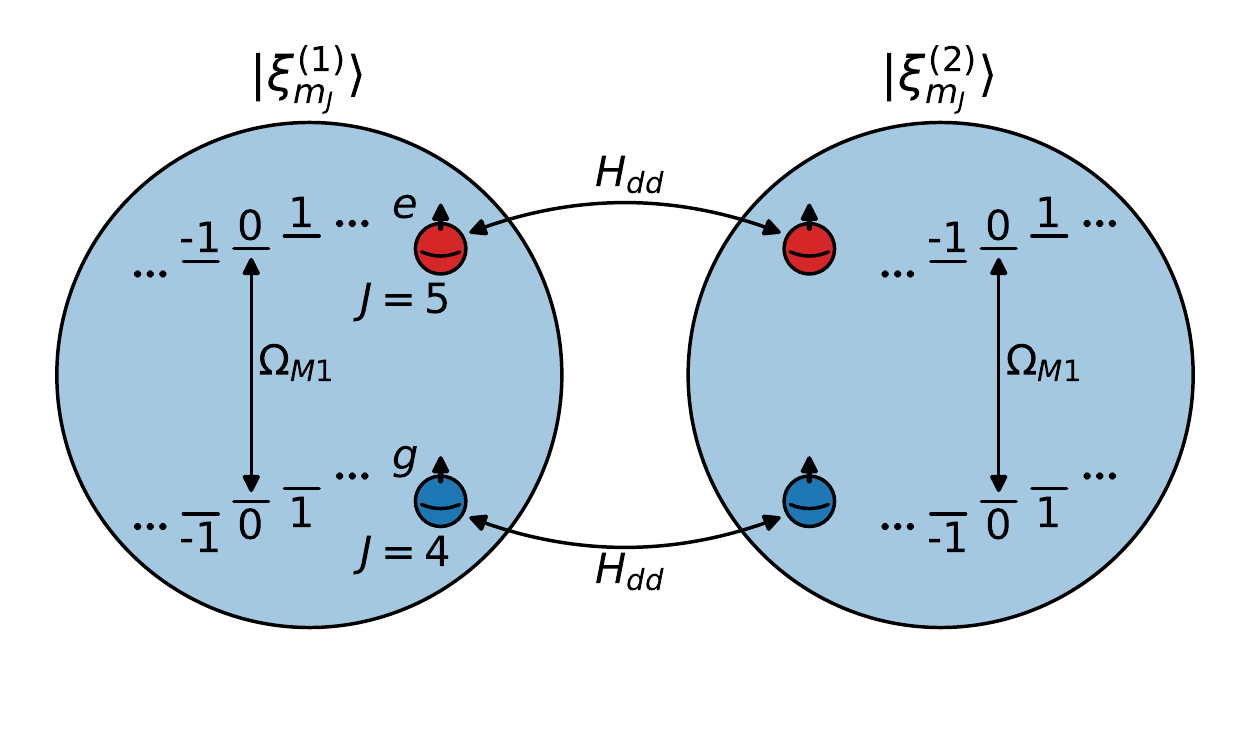}
    \caption{The secular dipole-dipole interaction of two atoms in the $a^3F_4$ (g) and $a^5F_5$ (e) clock states.  The $g$ and $e$ states have angular momentum quantum numbers $J=4$ and $J=5$, giving rise to $2J+1$ magnetic sublevels each. For brevity, the figure shows only three of these $m_J$ sublevels in each state, split by Zeeman shifts.  The clock drive $\Omega_{M1}$ couples the $|g_{m_J=0}\rangle$ and $|e_{m_J=0}\rangle$ states, and $H_{dd}$ couples the subspaces of the two atoms.}
    \label{fig:dd-cartoon}
\end{figure}

\begin{figure}[t]
    \centering
    \includegraphics[width=1\linewidth, keepaspectratio]{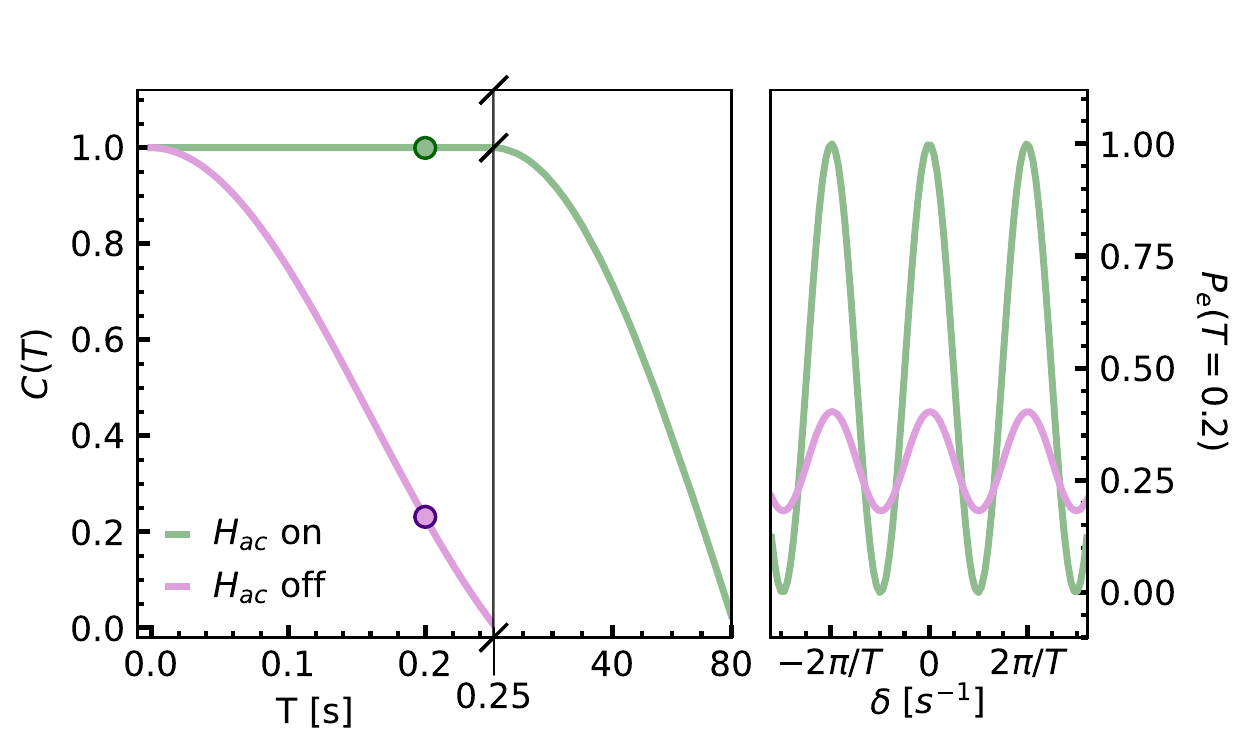}
    \caption{(left) Ramsey contrast of the clock transition as a function of the free evolution time $T$.  The contrast, shown in violet (green), dies quickly (slowly) when the ac Stark shift, $H_{ac}$, is excluded (included) from the simulation. Note the split in the $T$ axis and change in scale at 0.25 s.  (right) The resulting Ramsey fringes for $T=0.2$ s (indicated by circles in the left plot). We plot the probability an atom is found in state $|e_{0}\rangle$ as a function of the detuning $\delta$ from the clock transition..}
    \label{fig:dd-simulation}
\end{figure}

\begin{figure}[t]
    \centering
    \includegraphics[width=1\linewidth, keepaspectratio]{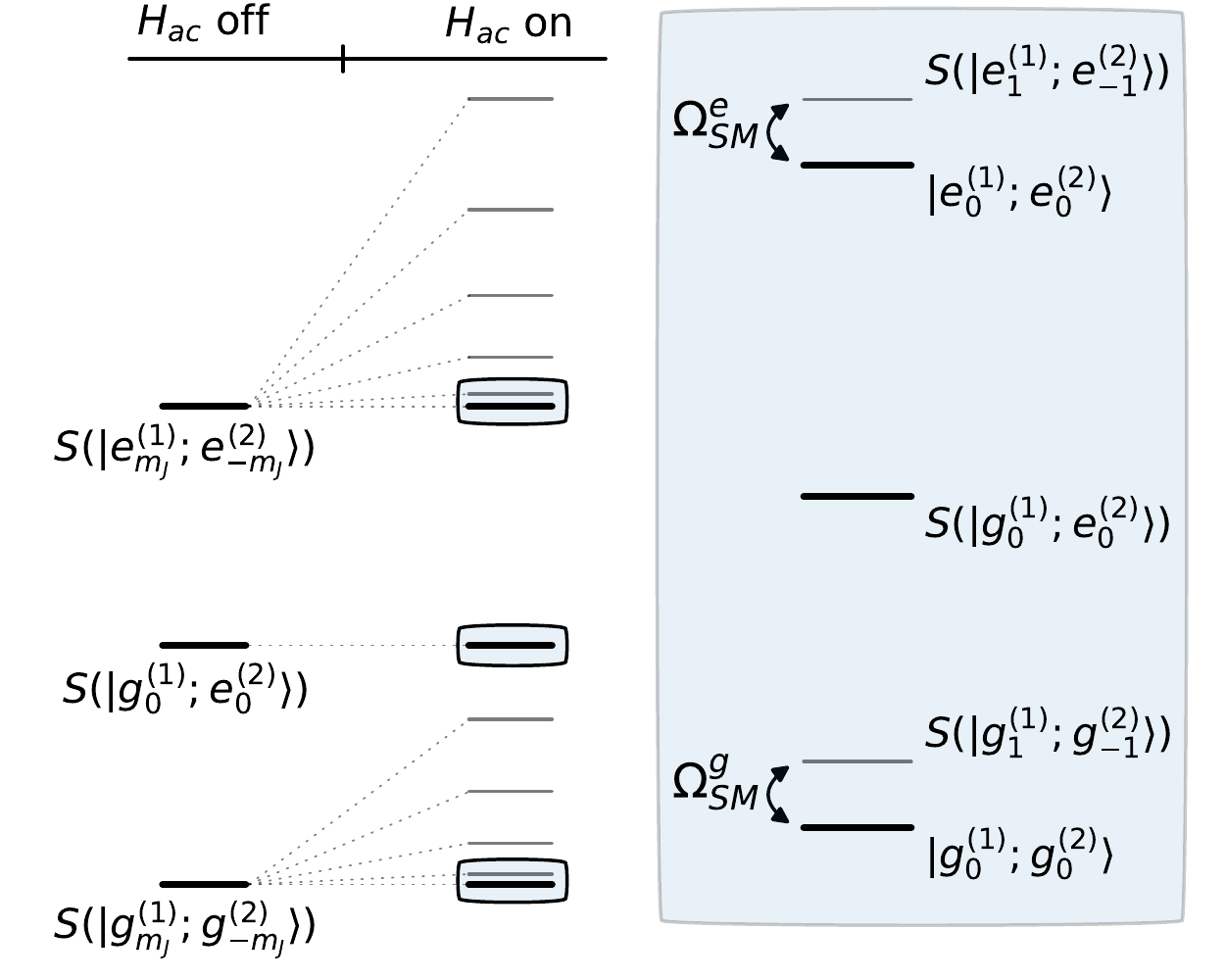}
    \caption{(left) The 2-atom level structure with and without the ac Stark shifts applied ($H_{ac}$).  Only states that are symmetric under particle exchange are shown, as these are the only states involved in the dynamics. The states $S(|g^{(1)}_{m_J\neq0};e^{(2)}_{-m_J}\rangle)$ are omitted because they are split from the $S(|g^{(1)}_{0};e^{(2)}_{0}\rangle)$ state by the differential Zeeman shift.  The dotted lines show that the degeneracy of the $m^{tot}_J=0$ sublevels is lifted when $H_{ac}$ is turned on.  The five boxed levels include the clock states and the states most strongly coupled to them by $H_{dd}$.  (right) The simplified picture, including only the levels boxed in the left panel.  $H_{dd}$ couples the states $|\xi^{(1)}_0;\xi^{(2)}_0\rangle$ and $S(|\xi^{(1)}_{m_J};\xi^{(2)}_{-m_J}\rangle)$ with rate $\Omega^{\xi}_{SM}$.}
    \label{fig:dd-protection-cartoon}
\end{figure}

To study the suppression of spin mixing, we simulate the secular dipole-dipole interaction of two nearest neighbor atoms in an optical lattice with and without accounting for the lattice-induced ac Stark shift.  Figure~\ref{fig:dd-cartoon} shows a schematic of the system under consideration.  We notate the single atom states by $|\xi^{(i)}_{m_J}\rangle$ where $\xi \in \{g,e\}$ refers to the upper or lower clock states, the superscript denotes the $i^{th}$atom, and the $m_J$ subscript is the magnetic sublevel.  We use $g$ to refer to the lower $a^3F_4$ clock state, and $e$ to refer to the upper $a^5F_5$ state.  The two atom system consists of $[(2J_g+1)+(2J_e+1)]^2=400$ states.

We simulate a simple Ramsey interferometry sequence which includes the dynamics of $H_{dd}$, the Zeeman Hamiltonian $H_Z$, and, optionally, the ac Stark Hamiltonian $H_{ac}$.  We initialize two atoms in the $|g^{(1)}_{0};g^{(2)}_0\rangle$ state, apply a $\pi/2$ pulse to each atom on the $g_0\rightarrow e_0$ clock transition, allow the two-atom state to evolve for a time $T$ under $H_Z$, $H_{dd}$, and (optionally) $H_{ac}$; apply a second $\pi/2$ pulse to each atom; and then determine the probability of an atom being in the $e_{0}$ state.  We vary the detuning of the clock drive frequency, $\delta$, to generate a Ramsey interferometry fringe.  From this fringe, we obtain the contrast $C(T)=(\max P_e(T, \delta) - \min P_e(T, \delta))$.

Figure~\ref{fig:dd-simulation} shows the results of the simulation.  The left panel demonstrates the decay of the Ramsey contrast with and without $H_{ac}$ included.  We break the axis at 0.25 s to illustrate that the tensor ac Stark shift extends the decay time of the contrast significantly.  Because we only simulate unitary dynamics of two neighboring atoms, revivals of the contrast are observed in the simulation beyond the times plotted in Figure~\ref{fig:dd-simulation}, but these would not occur in a true many-body situation as the coherence would spread between many particles and be lost.  The right panel shows the resulting Ramsey fringes taken at the time highlighted on the left side of the figure.  When no optical lattice is applied, the coherence on the clock transition quickly vanishes, leading to a loss of the Ramsey signal.  The fringe survives for $\sim85$ s when the lattice beams are on.

Figure~\ref{fig:dd-protection-cartoon} gives a simplified picture of the level structure at play.  $S(|\psi\rangle)$ is defined as the function that symmetrizes a multiparticle state by adding states with swapped $m_J$ and, if necessary, states in which the excited atom is switched.  The initial state of the Ramsey sequence ($|g^{(1)}_{0};g^{(2)}_0\rangle$) is symmetric under particle exchange and the Hamiltonian commutes with the exchange operator, so all states involved in the dynamics must remain symmetric.  The left panel of Figure~\ref{fig:dd-protection-cartoon} illustrates that in the absence of a tensor light field or dipole-dipole shifts, the $|\xi^{(1)}_{0};\xi^{(2)}_0\rangle$ states are degenerate with all the symmetrized states $S(|\xi^{(1)}_{m_J};\xi^{(2)}_{-m_J}\rangle) = 1/\sqrt2(|\xi^{(1)}_{m_J};\xi^{(2)}_{-m_J}\rangle + |\xi^{(1)}_{-m_J};\xi^{(2)}_{m_J}\rangle)$.  On the other hand, the figure shows the singly excited state $S(|g^{(1)}_{0};e^{(2)}_{0}\rangle)=1/\sqrt2(|g^{(1)}_{0};e^{(2)}_{0}\rangle + |e^{(1)}_{0};g^{(2)}_{0}\rangle)$ is not degenerate with states of mixed angular momentum (e.g.\ $S(|g^{(1)}_{1};e^{(2)}_{-1}\rangle)$) because the $g$ and $e$ states have different Zeeman splittings.

The dotted lines show the lifting of the degeneracy of the spin-mixed states by the ac Stark shift ($H_{ac}$). The splitting leads to the simplified level structure shown in the right panel of Figure \ref{fig:dd-protection-cartoon}. After the splitting, only the $S(|\xi^{(1)}_{1};\xi^{(2)}_{-1}\rangle)$ states remain energetically nearby the $|\xi^{(1)}_{0};\xi^{(2)}_{0}\rangle$ clock states.  The splitting between the $|\xi^{(1)}_{0};\xi^{(2)}_{0}\rangle$ and $S(|\xi^{(1)}_{1};\xi^{(2)}_{-1}\rangle)$ states by the tensor ac light shift is denoted $\Delta E_{tens}^\xi$.  With the parameters previously used to describe the optical lattice, this splitting is $h\times4(2)$ kHz for $gg$ atoms, $h\times4.8(6)$ kHz for $ee$ atoms.  The spin mixing term in $H_{dd}$ couples the nearby states with rates $\Omega^{\xi}_{SM}$ ($h \times 2.4$ Hz for $gg$ atoms, $h\times4.6$ Hz for $ee$ atoms). The coupling produces first-order level mixing at the $\Omega^{\xi}_{SM}/\Delta E^{\xi}_{tens}\sim10^{-3}$ level between these states --- small enough to not directly impact the coherence of the clock.

However at second-order, both of the clock states will experience an energy shift of $\Delta E_{\xi\xi}\sim-(\Omega^\xi_{SM})^2/\Delta E^\xi_{tens}$.  It is clear that the difference between these two shifts leads to a shift of the clock frequency, which from the simplified treatment is $\sim3$ mHz for two atoms. The sum of the shifts to the clock states also leads to decoherence through entanglement generation.  To understand this effect, we consider the state of the two atom system after it has evolved for a time $t$ during the Ramsey sequence:
\begin{align}
|\psi(t)\rangle = \frac{1}{2}\big[e^{-i\Delta E_{gg}t/\hbar}&|g_0^{(1)};g_0^{(2)}\rangle + e^{-i\Delta E_{ee}t/\hbar}|e_0^{(1)};e_0^{(2)}\rangle \nonumber \\
&+ |g_0^{(1)};e_0^{(2)}\rangle
+ |e_0^{(1)};g_0^{(2)}\rangle\big]
\end{align}
where we've included only the non-trivial phases accumulated from the second-order dipole-dipole interaction. If we trace over the second atom state space, we find the single atom reduced density matrix is:
\begin{align}
\rho^{(1)}(t) = &|g^{(1)}_0\rangle\langle g^{(1)}_0|+|e^{(1)}_0\rangle\langle e^{(1)}_0| \nonumber \\
&+ \big[(1 + e^{-i(\Delta E_{gg}+\Delta E_{ee})t/\hbar})|g^{(1)}_0\rangle\langle e^{(1)}_0| + h.c.\big]
\end{align}

From this, we can see the coherence is fully lost when $(\Delta E_{gg}+\Delta E_{ee})t/\hbar=\pm\pi$. Given the second-order dipole-dipole shifts calculated above, this corresponds to a full loss of coherence at $t\sim86$ s. This is in agreement with the lifetime seen in the simulation from Figure~\ref{fig:dd-simulation}.

\bibliography{ti_clock_refs}